# Cell processor implementation of a MILC lattice QCD application

**Guochun Shi**[1]

*National Center for Supercomputing Applications, University of Illinois*
*Urbana, IL 61801, USA*
E-mail: `gshi@ncsa.uiuc.edu`

**Volodymyr Kindratenko**

*National Center for Supercomputing Applications, University of Illinois*
*Urbana, IL 61801, USA*
E-mail: `kindr@ncsa.uiuc.edu`

**Steven Gottlieb**

*Department of Physics, Indiana University*
*Bloomington, IN 47405, USA*
E-mail: `sg@indiana.edu`

We present results of the implementation of one MILC lattice QCD application—simulation with dynamical clover fermions using the hybrid-molecular dynamics $R$ algorithm—on the Cell Broadband Engine processor. Fifty-four individual computational kernels responsible for 98.8% of the overall execution time were ported to the Cell's Synergistic Processing Elements (SPEs). The remaining application framework, including MPI-based distributed code execution, was left to the Cell's PowerPC processor. We observe that we only infrequently achieve more than 10 GFLOPS with any of the kernels, which is just over 4% of the Cell's peak performance. At the same time, many of the kernels are sustaining a bandwidth close to 20 GB/s, which is 78% of the Cell's peak. This indicates that the application performance is limited by the bandwidth between the main memory and the SPEs. In spite of this limitation, speedups of 8.7× (for 8×8×16×16 lattice) and 9.6× (for 16×16×16×16 lattice) were achieved when comparing a 3.2 GHz Cell processor to a single core of a 2.33 GHz Intel Xeon processor. When comparing the code scaled up to execute on a dual-Cell blade and a quad-core dual-chip Intel Xeon blade, the speedups are 1.5× (8×8×16×16 lattice) and 4.1× (16×16×16×16 lattice).

---

[1] Speaker

## 1. Introduction

Lattice Quantum Chromodynamics (QCD) calculations, used to simulate four-dimensional SU(3) lattice gauge theory, are very demanding. As a result, multiple attempts have been made to design and build chips and computers specific to QCD applications with a goal of high efficiency and high total performance. Measures of efficiency include CPU utilization, cost and power consumption. Examples of purpose built computers include digital signal processor based machines (QCDSP) [1], quantum chromodynamics on chip (QCDOC) [1], the CP-PACS computer [2], and the APE family of parallel computers [3]. With recent advances in more general-purpose parallel processors, such as dual- and quad-core CPUs, the Cell Broadband Engine™ (Cell/B.E.), and graphical processors (GPUs), researchers are also turning their attention to systems based on these chips as an alternative to special-purpose computers. The Cell/B.E., in particular, has attracted attention from the QCD community as it has been shown in [4] that "design parameters of the Cell/B.E. processor are remarkably close to the design space of the lattice QCD computations." A literature survey reveals, however, that most of the work on implementing a lattice QCD code on one of these processors is concerned with either a theoretical performance model, as in [4], [5], and [6], or an implementation of a subset of operations extensively used in lattice QCD codes, as in [6], [7], [8], [9], and [10].

The MILC [11] code is a publicly available lattice QCD code that is part of SPEC MPI benchmarks. In this work, we investigate the advantages and challenges of using the Cell/B.E. processor for running an entire MILC application. While several applications are included in the standard MILC distribution, in this initial study we consider simulations with dynamical clover fermions (clover_dynamical) using the hybrid-molecular dynamics *R* algorithm [12] (su3_rmd) as implemented in MILC version 7.4.

## 2. Target hardware

The Cell/B.E. system used in this study is a dual-Cell blade running in the IBM BladeCenter QS20 server [13] with the Cell/B.E. processor frequency of 3.2 GHz. The system runs Fedora Core 7 Linux OS, kernel 2.6.22, and Cell SDK 3.0 [14].

The Cell/B.E. processor is a heterogeneous system consisting of one 64-bit PowerPC® core called the Power Processor Element (PPE), eight Synergistic Processor Elements (SPEs), system memory, and I/O controller [15]. The processing elements are linked by an internal high-speed Element Interconnect Bus (EIB). The PPE is a 64-bit Power-Architecture-compliant core with 32KB first-level (L1) instruction and data caches and a 512-KB second-level (L2) cache. Its design is simplified in comparison with other four-issue out-of-order processors: it is a dual-issue, in-order execution design, two-way SMT processor. It can perform two double-precision or eight single-precision operations per clock cycle.

Each SPE consists of a Synergistic Processor Unit (SPU) and a Memory Flow Controller (MFC), which includes a DMA controller, a Memory Management Unit (MMU), a bus interface, and an atomic unit for synchronization with other SPEs and PPE. SPE is a Single Instruction, Multiple Data (SIMD) processor whose load and store instructions are performed in local address space only. The local address space is untranslated, unguarded, and non-coherent with respect to the system address space and is serviced by Local Storage (LS). The LS is full-pipelined, single-ported, 256KB SRAM that supports quadwords (16 bytes) and line (128 bytes) access. SPE cannot access main memory directly, but it can issue DMA operations to bring data from system memory to local storage or write data back to the system memory. SPE can perform eight single-precision floating-point operations in a single clock, or can issue four double-precision floating-point operations once every seven clock cycles.

For the 3.2 GHZ Cell/B.E., the EIB is capable of providing peak aggregate bandwidth of 204.8 GB/s. The memory interface controller provides 25.6 GB/s to system memory. The I/O controller provides peak bandwidth of 25 GB/s inbound and 35 GB/s outbound. When combining the PPE and eight SPEs, the 3.2 GHz Cell/B.E. has theoretical peak performance of 230.4 GFLOPS in single precision or 21.03 GFLOPS in double precision.

## 3. MILC implementation on Cell/B.E. processor

The MILC code is structured to enable an efficient parallel execution on multiprocessor systems using MPI. Its body consists of many small compute loops (kernels) that iterate over subsets of the 4D space-time lattice, and MPI scatter/gather operations in between. This structure provides the scalability necessary to efficiently execute the application on a large distributed memory system. However, as a result, no single kernel is responsible for more than ~20% of the overall execution time.

When porting the code to the Cell processor, our main design goal was to preserve MILC's ability to scale to a large number of compute nodes. This dictated the overall implementation approach: keep the MPI-based distributed execution framework intact (to be executed on the Cell's PPE) while accelerating individual kernels on the Cell's SPEs.

The MILC application core consists of 27 major subroutines. While many of the subroutines are responsible for only a small fraction of the overall execution time on the 2.33 GHz Intel Xeon system, their run time increases on average by about a factor of 2 when executed on the Cell's PPE. Therefore they all need to be ported to the Cell/B.E. SPEs in order to avoid introducing additional computational overhead on the PPE. Since many of these subroutines consist of multiple compute kernels with the MPI scatter/gather operations in between, we identified 54 unique kernels for implementation on the SPEs.

In case of the 8×8×16×16 lattice model, the 54 kernels of interest are called 13,408 times. It is impractical to spawn a new SPE thread each time a new kernel is executed because the SPE thread execution overhead will have an adverse impact on the overall application performance. Fortunately, the kernels are small in terms of the actual lines of code, so they can be bundled in a single library of SPE-resident subroutines. Since individual subroutines residing on the SPE cannot be directly accessed from the PPE code, we implemented a thin interface running on the SPE as the main SPE thread. Its only function is to invoke an appropriate SPE-resident library subroutine. Thus, only one thread per SPE is invoked at the start of the application.

Each compute kernel in the original CPU-based code is replaced with a small wrapper subroutine, executed on the PPE, that i) sets up the task structures specific for each individual kernel, ii) notifies the SPEs via mailboxes, and iii) waits for the completion message from all the SPEs. The task structure created by these subroutines is copied to a container padded to a multiple of 128 bytes, and the pointer to the container is sent to the SPE as a mailbox message. Each SPE upon receiving a mailbox message converts the message to a pointer in the global memory space and fetches the first eight bytes via DMA. The first four bytes provide the unique SPE-resident subroutine identifier and the next four bytes indicate the kernel-specific task structure size. The SPE then transfers the entire kernel-specific structure to its LS and passes the control to the corresponding SPE-resident subroutine. Once the calculations are done, results are transferred back to the main memory and a mailbox message is sent to the PPE indicating the completion of the SPE task.

Performance of most of the kernels is bound by the memory-to-local store bandwidth, as will be shown later. Therefore, in porting each individual kernel, special care is taken to maximize the data transfer bandwidth and overlap the calculations with DMA transfer calls in addition to the usual vectorization of the code for SPE's SIMD engine. For any given kernel, there are three types of input data: elements in a lattice site, elements in contiguous memory (usually a temporary memory region created inside MILC for temporary use), and elements in the neighboring site (neighbors in term of (x, y, z, t); they are not physically adjacent to each other in memory). There are only two types of output data: elements in a lattice site and elements in contiguous memory. A common DMA engine is written to load input data into SPEs' LS and output data back to the main memory.

The Cell/B.E. processor delivers the best memory-to-local store bandwidth when both source address and destination address are aligned at the 128-bytes boundaries and the amount of data to be transferred is a multiple of 128 bytes. Non-aligned DMA requests run at the half of the bandwidth due to the fact that two bus requests instead of one are needed for each cache line of data. We tried several approaches to deal with the data alignment issue and selected the memory padding approach. The lattice is allocated to be aligned at 128 bytes and the most commonly used structures are padded to 128 bytes or multiples of that. Two data structures are padded for this reason: su3_matrix is changed from 3x3 matrix to 4x4 matrix, thus changing the size of su3_matrix from 72 bytes to 128 bytes, and su3_vector is changed from a vector of three complex variables to a vector of four complex variables. This change also makes one of the other commonly used data structures, fwilson_vector, 128 bytes. The obvious disadvantage of this approach is that more bytes of data have to be transferred between the main memory and SPEs' LS. However, padding helps both to better use the bandwidth between main memory and local storage and to make writing SIMD instructions easier since the data structures are already aligned at 16-byte boundaries. For these same reasons, we also pad the lattice site data with a few additional bytes to be a multiple of 128 bytes in length. Data from each site is usually accessed in a strided manner, therefore padding it to multiples of 128 bytes helps to ensure better bandwidth utilization. However, simply padding it to multiples of 128 bytes is not sufficient to achieve the full memory-to-SPE's-LS bandwidth. Main memory attached to each Cell/B.E. processor in the Cell blade consists of 16 banks distributed by cacheline address (128 B) with address 0 in bank 0, address 128 in bank 1, etc. In aggregate, the banks can sustain a data transfer rate of 25.6 GB/s. However, for strided access, depending on the stride size, all banks or only some banks may be used. Any time the stride size has common factor with 16, some banks are not used and the full memory-to-LS bandwidth cannot be achieved. Thus, with an odd stride size we achieve the maximum bandwidth of 25.38 GB/s, whereas with the stride size of 16 we achieve only 2.13 GB/s because only one out of 16 memory banks is used all the time. Therefore, we pad the lattice site data to the nearest odd stride size times 128 bytes, which is 21×128 bytes in our case, to ensure the full memory-to-LS bandwidth utilization.

**4. Results and discussion**

The 54 computational kernels that we ported to the Cell/B.E. processor are responsible for 98.8% of the overall execution time on the 2.3 GHz Intel Xeon chip; only 1.2% of the overall execution time is due to the remaining code (Fig. 1). However, once ported to the Cell/B.E. processor, runtime of the remaining code increases to more than 30% of the overall execution time. We observe that i) execution time of the part of the code that remains on the Cell's PPE slows down more than three times as compared to the execution time on the single core of the Intel Xeon chip, and ii) the part of the code ported to the eight SPEs speeds up more than 12 times as compared to the Intel Xeon execution time. These observations hold true for both lattice sizes tested in this work. It is clear that PPE becomes the bottleneck in achieving any substantial performance increase beyond this point. As we have shown, performance of the code executed on the PPE is limited by the main RAM-to-PPE bandwidth.

The QS20 IBM Cell blade consists of two Cell/B.E. processors mounted on the same board with a high-speed coherent interface between the two chips running at 20 GB/s in each direction. We investigate two ways to scale up the application on the dual-chip board:

1) Run the application on one PPE while offloading the 54 kernels to 16 SPEs (NUMA bar in Fig. 1). Since the performance is largely determined by the bandwidth between main memory and the SPEs, we allocate memory alternatively among two Cell processors to maximize the bandwidth. In this case, we observe a slight increase in the time spent on the PPE and a small decrease of the time spent on the SPEs, with overall runtime decreasing 18% to 22% for different lattice size.

2) Run the application on two PPEs as two MPI processes with each MPI process computing half of the full grid size (MPI bar in Fig. 1). Each process runs on one PPE and offloads the computational kernels to its own eight SPEs. The two processes communicate using MPI built on top of the shared memory. While the runtime for the SPE part of the code decreases, we observe that the runtime of the remaining PPE part of the code increases to well over 50% of the overall execution time. Further profiling shows the added MPI communication overhead is over 70%. The overall performance of this implementation slightly

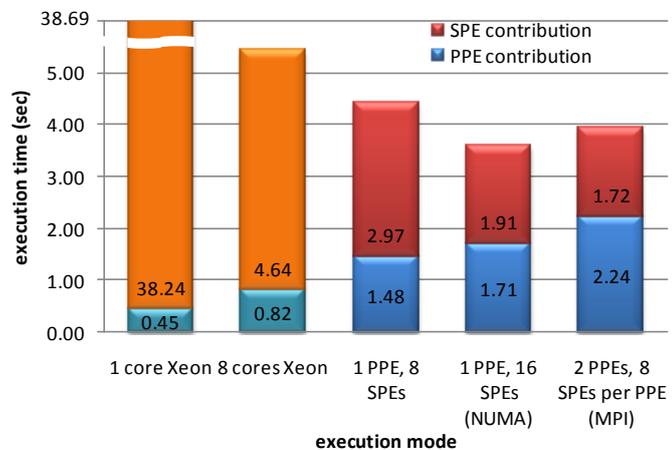

**Figure 1.** Execution time of the MILC Cell/B.E. implementation on the Cell blade for the lattice size of 8x8x16x16. Single- and multi-core Intel Xeon performance is provided for reference.

increases when comparing to the first execution schema.

These results are as expected: By going from eight SPEs to 16 SPEs, we increased the number of compute engines by a factor of two, allowing the SPE-resident code to run faster by making available more memory bandwidth. However, by spreading data between the memories of two Cell/B.E. processors, we did not increase the effective memory-to-SPEs local storage bandwidth by a factor of two because in many instances SPEs from one Cell/B.E. chip access memory attached to the other Cell/B.E. chip, resulting in a higher latency and a lower bandwidth. An MPI implementation makes a better data localization per Cell/B.E. processor, but it is still not ideal. However, MPI itself runs quite slowly on the PPE, thus reducing the overall performance.

As we mentioned earlier, performance of most of the kernels running on the SPEs is bounded by the memory-to-local store bandwidth. We observe that we only infrequently achieve more than 10 GFLOPS with any of the kernels, which is just over 4% of the combined Cell PPEs' peak performance. At the same time, only five kernels achieve less than 10 GB/s memory bandwidth utilization, and many of the kernels are sustaining a bandwidth close to 20 GB/s, which is 78% of the peak.

The Cell/B.E. blade system available in our lab consists of two blades interconnected with Gigabit Ethernet, and we were able to run the code across two Cell/B.E. blades using MPI. While execution time of the Cell/B.E. SPE code increases proportionally to the dataset size, we observe a disproportionate increase in the execution time of the PPE code, largely due to the MPI-related subroutines. We also observe that the MPI code execution time varies from one run to another by as much as 50%, a phenomenon that we yet have to explain.

## 5. Conclusions

In summary, we took an existing production-grade lattice QCD code, ported it to Cell/B.E. processor and achieved speedups of 8.7× (for 8×8×16×16 lattice size) and 9.6× (16×16×16×16 lattice size) compared to a single-core Intel Xeon processor. When we scaled up the code to run on minimal size compute systems (a quad-core dual-chip Intel Xeon blade and a dual-Cell/B.E. blade), the achieved speedups were 1.5× and 4.1×, respectively. The MILC code runs faster on the Cell/B.E. blade and the speedup depends on the lattice size and the number of processor elements used. By changing lattice size from 8×8×16×16 to 16×16×16×16, we effectively quadrupled the number of calculations and amount of data to be processed. As a result, the overall execution time increased only by a factor of 3.8. This superlinear speedup is due to better bandwidth utilization by the SPEs for larger datasets. The overhead associated with DMA transfers is better amortized when there is more data to transfer. Thus, for 16 SPEs it takes only 3.6 times longer to process a dataset that is four times larger. However, the remaining PPE code takes four times longer to execute.

Finally, we note that our results differ from the model predictions reported in some of the earlier work referenced in Section 1. We have not considered modifying the structure of the application to suit the Cell/B.E. architecture, as was the case in many of the referenced publications. We considered only accelerating some of its parts and our results clearly show that with this approach we can achieve only a limited performance improvement over the multicore platform. Achieving more substantial performance improvements will require significant code redesign. We are interested in exploring that task.

## 6. Acknowledgement

This work was funded by National Science Foundation grant SCI 05-25308 and by the IBM Linux Technology Center and in part is based on the MILC collaboration's public lattice gauge theory code (http://physics.utah.edu/~detar/milc.html).